\documentclass[lettersize,journal]{IEEEtran}
\usepackage{array}
\usepackage{amsfonts}
\usepackage{amsbsy}
\usepackage{amssymb}
\usepackage{times}
\usepackage{graphicx}
\usepackage{enumerate}
\usepackage[usenames]{color}
\usepackage[dvips]{pstcol}
\usepackage{epstopdf}
\usepackage[caption=false,font=normalsize,labelfont=sf,textfont=sf]{subfig}
\usepackage{cite}
\usepackage{amssymb}
\usepackage{amsfonts}
\usepackage{graphicx}
\usepackage{epsfig}
\usepackage{psfrag}
\usepackage{xcolor}
\usepackage{amsfonts, bm}
\usepackage{epstopdf}
\usepackage{cite}
\usepackage{color}
\usepackage{xcolor}
\usepackage{subfig}
\usepackage{verbatim}
\usepackage{multirow}
\usepackage{array}
\usepackage{booktabs}
\usepackage{amsthm}
\usepackage{makecell}
\usepackage{hyperref}

\usepackage[linesnumbered, ruled]{algorithm2e}
\usepackage{algpseudocode}
\usepackage{amsmath}

\IEEEoverridecommandlockouts

\begin{document}

	\title{Rate-Adaptive Generative Semantic Communication Using Conditional Diffusion Models}
	\author{Pujing Yang, \textit{Graduate Student Member, IEEE}, Guangyi Zhang, \textit{Graduate Student Member, IEEE}, \\
		and Yunlong Cai, \textit{Senior Member, IEEE}
		\thanks{
			This work was supported in part by the  Major Key Project of Peng Cheng Laboratory under Grant PCL2023AS1-2,  in part by the National Natural Science Foundation of China under Grant U22A2004, and in part by Zhejiang Provincial Key Laboratory of Information Processing, Communication and Networking (IPCAN), Hangzhou 310027, China. (Corresponding author: Yunlong Cai)
			
			P. Yang and G. Zhang are with the College of Information Science and Electronic Engineering, Zhejiang University, Hangzhou 310027, China (e-mail: yangpujing@zju.edu.cn; zhangguangyi@zju.edu.cn).
			
			Y. Cai is with the College of Information Science and Electronic Engineering, Zhejiang University, Hangzhou 310027, China, and also with the Peng Cheng Laboratory, Shenzhen 518071, China (e-mail: ylcai@zju.edu.cn).
			
			Our project can be found at \href{https://github.com/zhang-guangyi/cdm-jscc}{\textit{https://github.com/zhang-guangyi/cdm-jscc}}.
			
	} }

	\maketitle
	\vspace{-3.3em}
	\begin{abstract}
		Recent advances in deep learning-based joint source-channel coding (DJSCC) have shown promise for end-to-end semantic image transmission.
		However, most existing schemes primarily focus on optimizing pixel-wise metrics,  which often fail to align with human perception, leading to lower perceptual quality.
		In this letter, we propose a novel generative DJSCC approach using conditional diffusion models to enhance the perceptual quality of transmitted images. 
		Specifically, by utilizing entropy models, we effectively manage transmission bandwidth based on the estimated entropy of transmitted symbols. These symbols are then used at the receiver as conditional information to guide a conditional diffusion decoder in image reconstruction.
		Our model is built upon the emerging advanced mamba-like linear attention (MLLA) skeleton, which excels in image processing tasks while also offering fast inference speed. 
		Besides, we introduce a multi-stage training strategy to ensure the stability and improve the overall performance of the model.
		Simulation results demonstrate that our proposed method significantly outperforms existing approaches in terms of perceptual quality. 
		
		
		
	\end{abstract}
	
	
	\IEEEpeerreviewmaketitle
	
	\section{Introduction}

	The rapid development of sixth-generation (6G) communication systems has driven the rise of various smart applications, such as Virtual Reality (VR) and the Internet of Everything (IoE) \cite{you2024next, xie2021task, JSCC,transformeraided}. These services demand enhanced communication efficiency to manage the massive inflow of data traffic. In this context, semantic communications have emerged as a new paradigm,  attracting significant attention. Unlike traditional transmission systems that rely on separate source and channel coding, semantic communications focus on accurately transmitting the underlying semantic information of digital data. This approach integrates source and channel coding for joint optimization, a technique known as joint source-channel coding (JSCC).
	
	Recently, the integration of deep learning into wireless communication system designs has gained traction, driven by the exceptional information processing capabilities of various deep learning models. In this context, deep learning-based JSCC (DJSCC) has stimulated significant interest, particularly through the use of autoencoders (AEs) and their variants variational AEs. DJSCC maps input images into low-dimensional vectors for transmission. A pioneering work in this area is DeepJSCC\cite{JSCC}, which is built on an AE-based framework that allows for joint optimization and mitigates the cliff-edge effect commonly observed in traditional separation-based schemes. Moreover, the authors in \cite{transformeraided} introduced a Transformer-based framework that enhances image fidelity by incorporating channel feedback to adaptively reconstruct images under varying wireless conditions. Inspired by entropy model-based source coding \cite{balle2018variational}, the authors in \cite{NTSCC} proposed a rate-adaptive JSCC system, where the number of transmitted symbols is determined by the entropy estimated by the entropy models. This work was further evolved in \cite{MDJCM} into a digital version, enabling practical application in modern digital systems.

	Despite the superior performance of current DJSCC transmission systems, they are typically optimized for mean-square error (MSE)-based distortion metrics such as peak signal-to-noise ratio (PSNR), which assesses pixel-level similarity between reconstructed and source images. However, it is increasingly recognized that high pixel-level similarity does not necessarily indicate high perceptual quality, which reflects how humans perceive the image's realism and visual appeal \cite{blau2019rethinking}. 
	Generative models, including generative adversarial networks (GANs) and denoising diffusion probabilistic models (DDPM) \cite{ho2020DDPM}, excel at generating images with high perceptual quality from a predefined prior distribution, demonstrating their groundbreaking potential in generative semantic communications \cite{wang2022perceptual, cddm2024wu, diffgo, grassucci2023generative, jiang2024diffsc, diffusion2023high,  li2022domain}. 
	To improve the perceptual quality of transmitted images, the authors in \cite{wang2022perceptual} incorporated perceptual loss into a GAN-based DJSCC system. 
	In \cite{cddm2024wu}, the authors leveraged DDPM to mitigate channel noise in wireless communications.
	Furthermore, in [14], DJSCC was employed for initial reconstruction, which was then refined by DDPM to further improve image quality.
	Another approach in \cite{diffusion2023high} divided the reconstructed image into two components: range-space and null-space. 
	\color{black}
	The range-space that captures the primary structure is transmitted via DJSCC, while the null-space, refining details, is generated at the receiver using diffusion models.
	Although these methods achieved performance improvements, the decoding process at the receiver is time-consuming, as denoising requires hundreds of steps. Additionally, the training is performed module by module without joint optimization, potentially leading to performance loss. Besides, they typically support fixed-rate transmission, lacking the ability to determine optimal bandwidth for each image.
	
	
	In this letter, we introduce a novel framework called conditional diffusion models-based generative DJSCC (CDM-JSCC) for wireless image transmission.
	Unlike previous methods that rely on DDPM noise prediction (referred to as $\epsilon$-prediction)\cite{diffusion2023high}, our method utilizes $\mathcal{X}$-prediction \cite{CDC} to directly predict the source image. 
	$\mathcal{X}$-prediction offers comparable performance to $\epsilon$-prediction but with significantly fewer steps, as its objective function resembles an autoencoder loss, enabling data reconstruction in a single iteration.
	Moreover, we optimize transmission bandwidth by utilizing entropy models to estimate the entropy of transmitted symbols. At the receiver, the transmitted symbols are utilized as conditional information to guide the conditional diffusion decoder to reconstruct images.
	Recently, the mamba-like linear attention (MLLA)  technique has demonstrated superior performance in image processing tasks, benefiting from parallel computation and fast inference \cite{MLLA}. Building on this, we design our JSCC encoder and decoder with MLLA as the core architecture.
	Furthermore, we propose a multi-stage training strategy to achieve joint optimization, resulting in substantial enhancements in perceptual quality.
	
	\begin{figure}[t]
		\begin{centering}
			\includegraphics[width=0.40\textwidth]{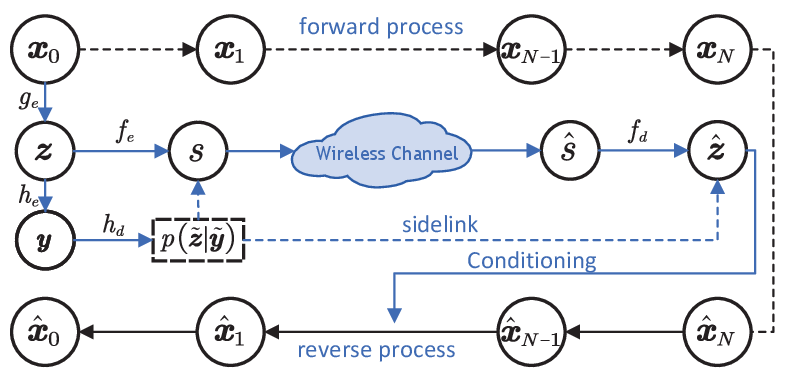}
			\par\end{centering}
		\caption{The overview of the proposed CDM-JSCC. The blue line represents the transmission process, and the black line denotes the diffusion process. For simplification, we ignore quantization in the figure.}
		\label{Frame}
	\end{figure}


	\section{Proposed CDM-JSCC Framework} \label{S2}
	In this section, we propose the framework of the proposed CDM-JSCC to realize rate-adaptive image transmission. To provide a comprehensive understanding, we begin by giving	an overview of CDM-JSCC and then proceed to detail the conditional diffusion models-based decoder.

	\subsection{Rate-Adaptive JSCC}
	As illustrated in Fig. \ref{Frame}, the source image, represented
	by a pixel intensity vector $\bm{x}_0 \in \mathbb{R}^{n}$, is first processed by a nonlinear transform coding encoder $g_e$, producing a low-dimensional latent representation $\bm{z} = g_e(\bm{x}; \bm{\theta}_g)$, where $\bm{\theta}_g$ encompasses the trainable parameters.
	Inspired by \cite{balle2018variational}, we estimate the distribution of $\bm{\tilde{z}}$, i.e., the quantized version of $\bm{z}$, by utilizing hyperprior entropy models. Specifically, we introduce an additional latent feature $\bm{y} = h_e(\bm{z}; \bm{\theta}_h)$, serving as side information to capture the dependencies among the elements in $\bm{\tilde{z}}$, where $h_e$ represents the parametric analysis transform, and $\bm{\theta}_h$ denotes its trainable parameters. 
	Each $\tilde{z}_i$ in $\bm{\tilde{z}}$ is variationally modeled as a Gaussian distribution with the standard deviation $\sigma_i$ and mean $\mu_i$ predicted based on the quantized $\bm{\tilde{y}}$ as:
	\begin{equation}
		p_{\bm{\tilde{z}} | \bm{\tilde{y}}}\left({\bm{\tilde{z}}} | \bm{\tilde{y}}\right)= \prod_i \left(\mathcal{N}(\!\tilde{\mu}_i, \tilde{\sigma}_i^2) * \mathcal{U}(-\frac{1}{2}, \frac{1}{2})\right) \left(\tilde{z}_i\right),
	\end{equation}
	where $(\bm{\tilde{\mu}}, \bm{\tilde{\sigma}})=h_s(\bm{\tilde{z}}; \bm{\phi}_h)$, $h_s$ is the parametric synthesis transform, $\bm{\phi}_h$ denotes its trainable parameters, and $*$ represents the convolutional operation. Moreover, we convolve each element with a standard uniform density $\mathcal{U}(-\frac{1}{2}, \frac{1}{2})$ to enable a better match of the prior to the distribution of $\bm{\tilde{z}}$ \cite{balle2018variational}.

	In this way, we are able to control the channel bandwidth ratio (CBR) for each image based on the estimated entropy of $\bm{\tilde{z}}$. Images with rich details, characterized by higher entropy, may require more symbols for transmission. \color{black} Specifically, $\bm{z}$ comprises multiple embedding vectors $z_i$ of length $C$. The JSCC encoder $f_e$ adjusts the length of each vector to ${k}_i=Q\left(-\beta \log p_{\tilde{z}_i | \boldsymbol{\tilde{y}}}\left(\tilde{z}_i | \boldsymbol{\tilde{y}}\right)\right)$, where $Q$ denotes the quantization operation, $\beta$ controls the relation between the prior $p_{\hat{z}_i | \boldsymbol{\tilde{y}}}\left(\hat{z}_i | \boldsymbol{\tilde{y}}\right)$, and the expected length of the corresponding symbol vector $s_i$. 	In particular, when the entropy of $\tilde{z}_i$ is high, a higher CBR is adopted, resulting in a larger $k_i$. This process is expressed as $
	\bm{s} = f_e(\bm{z}, p_{\tilde{\bm{z}} | \boldsymbol{\tilde{y}}}\left(\tilde{\bm{z}} | \boldsymbol{\tilde{y}}\right); \bm{\theta}_f) \in \mathbb{C}^{k},$ where $\bm{s}$ denotes the channel input symbols, $k$ is the number of transmitted symbols, and $\bm{\theta}_f$ denotes the trainable parameters.  Additionally, to meet the energy constraints of real-world communication systems, we ensure that $\bm{s}$ satisfies an average power constraint before transmission.

	Then, $\bm{s}$ is transmitted through a noisy wireless channel, which is denoted by $\eta$. As the additive white Gaussian noise (AWGN) channel is adopted in our work, this process follows $\bm{\hat{s}} \triangleq \eta(\bm{s}) =\bm{s} + \bm{n}$, where $\bm{n} \sim \mathcal{CN}(0,\sigma^2 \bm{I}_{k \times k})$ is a complex Gaussian vector with variance $\sigma^2$. 
	At the receiver, $\bm{\hat{s}}$ is first decoded to obtain the reconstructed latent $\bm{\hat{z}} = f_d(\bm{\hat{s}}; \bm{\phi}_f)$, where  $f_d$ represents the JSCC decoder, and $\bm{\phi}_f$ denotes its trainable parameters.
	The receiver utilizes $\bm{\hat{z}}$ as an additional “content” latent to guide the conditional diffusion decoder $g_d$ to reconstruct image $\bm{\hat{x}}_0 = g_d(\bm{\hat{x}}_N, \bm{\hat{z}}; \bm{\phi}_g)$, where $\bm{\hat{x}}_N$ is the randomly sampled Gaussian source and $\bm{\phi}_g$ represents the trainable parameters, as shown in Fig. \ref{Frame}. The details of the conditional diffusion decoder will be provided in the following section.
	
	\begin{figure*}[t]
		\begin{centering}
			\includegraphics[width=0.78\textwidth]{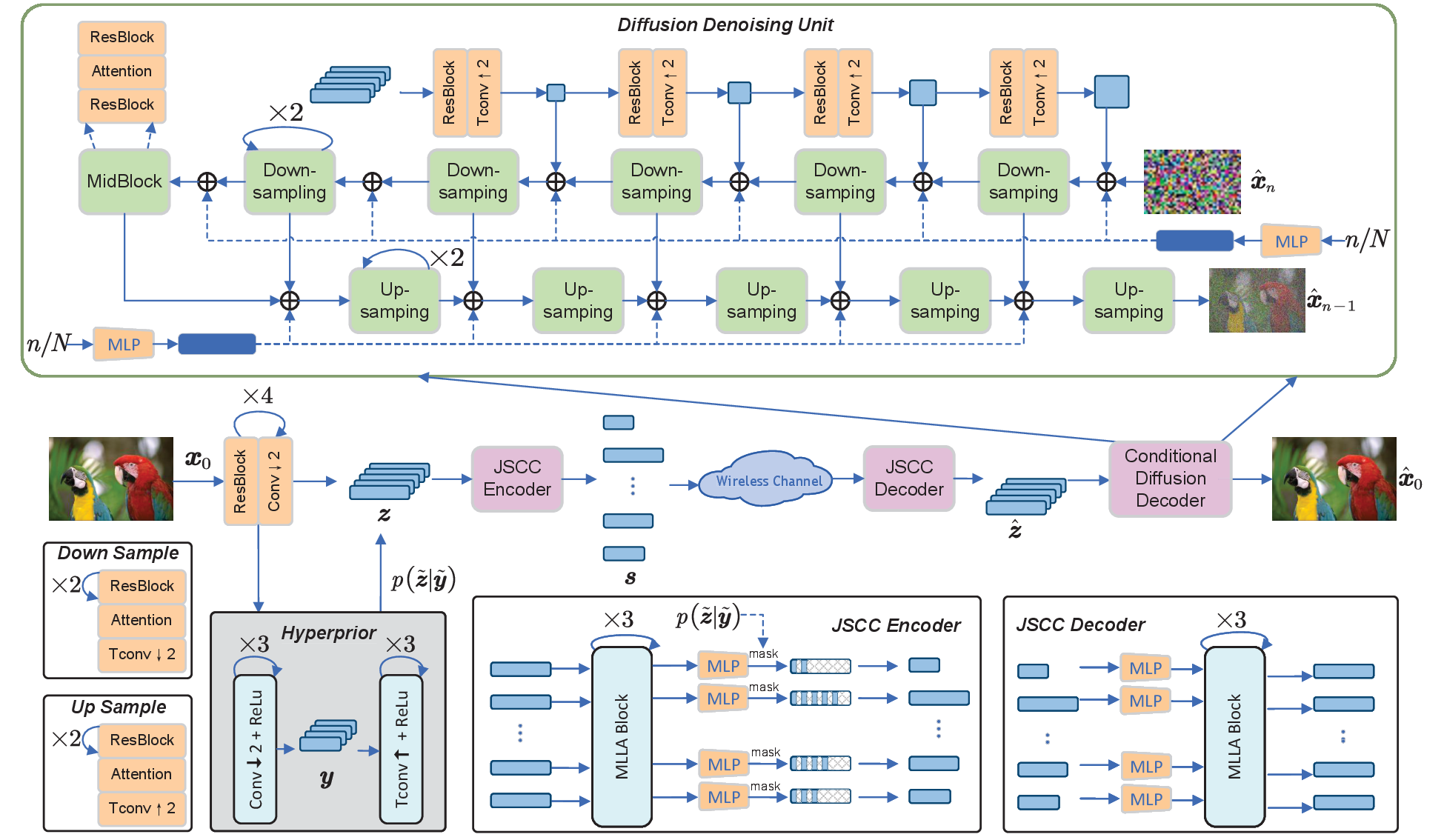}
			\par\end{centering}
		\caption{The architecture of the proposed system.}
		\label{Architecture}
	\end{figure*}
	
	\subsection{Conditional Diffusion Decoder} \label{condiff}	
	We build our decoder on conditional diffusion models for their significant success in generative tasks. 
	
	The core idea of diffusion models is to transform an image $\bm{x}_0$ into a Gaussian distribution by progressively adding noise to it, referred to as the forward process $q$, resulting in a sequence of increasingly noisy versions $\bm{x}_1, \bm{x}_2,..., \bm{x}_N$. Then, the reverse process $p_{\theta}$ generates high-quality samples by reversing this process.
	The two Markov processes at step $n$ can be respectively described as follows:
	\begin{equation}
		\label{eq:diffusion1}
		\begin{aligned}
			\quad q(\bm{x}_n|\bm{x}_{n-1}) = {\mathcal N}(\bm{x}_n; \sqrt{1-\beta_n}\bm{x}_{n-1}, \beta_n \bm{I}),
		\end{aligned}
	\end{equation}
	\begin{equation}
		\label{eq:diffusion}
		\begin{aligned}
			\quad p_\theta(\bm{x}_{n-1} | \bm{x}_n) = {\mathcal N}(\bm{x}_{n-1}; \mu_\theta(\bm{x}_n,n), \Sigma_\theta(\bm{x}_n, n)),
		\end{aligned}
	\end{equation}
	where the variance $\beta_n$ is held constant as hyperparameters, the reverse process mean $\mu_\theta(\bm{x}_n, n)$ is parameterized by a neural network, and the variance $\Sigma_\theta(\bm{x}_n, n)$ is always set to $\beta_n \bm{I}$.
	
	The diffusion models are typically trained to predict the accumulated noise $\epsilon$ that perturbs  $\bm{x}_0$ to $\bm{x}_n$, a process known as $\epsilon$-prediction, with the loss function being:
	\begin{equation}
		\label{eq:diffloss}
		\mathcal{L}(\theta, x_0)=\mathbb{E}_{\bm{x}_0, n,\epsilon}||\epsilon - \epsilon_\theta(\bm{x}_n, n)||^2,
	\end{equation}
	where $n \sim \text{Uniform}(1, ..., N)$, $\epsilon \sim \mathcal{N}(0, \bm{I})$,  $\bm{x}_n = \sqrt{\bar{\alpha}_n} \bm{x}_0 + \sqrt{1 - \bar{\alpha}_n} \epsilon$, and ${\bar{\alpha}_n=\prod_{i=1}^n (1-\beta_i)}$.

	In our conditional diffusion decoder, $\bm{\hat{z}}$ is taken as the condition for the reverse process. Consequently, the forward process remains unchanged as (\ref{eq:diffusion1}), and the reverse process is replaced by:
	\begin{equation}
		\label{eq:con_diffusion}
		\begin{aligned}
			\quad p_\theta(\bm{x}_{n-1} | \bm{x}_n, \bm{\hat{z}}) = {\cal N}(\bm{x}_{n-1}; \mu_\theta(\bm{x}_n, \bm{\hat{z}}, n), \beta_n{\bf I}).
		\end{aligned}
	\end{equation}
	
	Analogous to (\ref{eq:diffloss}), we modify the training objective as follows:
	\begin{equation}
		\label{eq:distortion_diffusion}
		\begin{aligned}
			\mathcal{L}(\theta, \bm{x}_0) 
			= \mathbb{E}_{n, \epsilon}||\bm{x}_0 - \mathcal{X}_\theta(\bm{x}_n, \bm{\hat{z}}, \frac{n}{N})||^2.
		\end{aligned}
	\end{equation}	
	First, we employ $\mathcal{X}$-prediction \cite{CDC} to predict the source image instead of using $\epsilon$-prediction. $\mathcal{X}$-prediction requires only a few denoising steps yet achieves performance comparable to $\epsilon$-prediction, which involves hundreds of denoising steps. This is because the optimization objective resembles an autoencoder loss, allowing the model to reconstruct the source image in a single iteration. Moreover, we replace the step $n$ with a normalized step $\frac{n}{N}$, enabling the use of a smaller $N$ during testing compared to training, thereby accelerating the decoding process.
	
	During inference, images are generated using ancestral sampling with Langevin dynamics as follows:
	\begin{equation}
		\label{eq:111}
		\begin{aligned}
			\bm{x}_{n-1}& = \sqrt{\bar{\alpha}_n} \mathcal{X}_\theta(\bm{x}_n, \bm{\hat{z}}, \frac{n}{N}) - \sqrt{1-\bar{\alpha}_n} \epsilon_\theta(\bm{x}_n,n,\frac{n}{N}), \\
			&\epsilon_\theta(\bm{x}_n, \bm{\hat{z}}, \frac{n}{N})=\frac{\bm{x}_n-\sqrt{\bar{\alpha}_n}\mathcal{X}_\theta(\bm{x}_n, \bm{\hat{z}}, \frac{n}{N})}{\sqrt{1-\bar{\alpha}_n}}.
		\end{aligned}
	\end{equation}
	
	\section{Training Strategy and Model Architecture} \label{S3}
	In this section, we first present the architecture of CDM-JSCC. Then, we analyze the proposed loss function. Finally, we propose a multi-stage training algorithm to ensure the stability and improve the overall performance.
	\subsection{Model Architecture}
	As shown in Fig. \ref{Architecture}, the proposed model consists of a transmitter, a wireless channel, and a receiver. We adopt the AWGN channel in our work, which is incorporated as a non-trainable layer in the architecture.

	The transmitter comprises an entropy encoder and a rate-adaptive JSCC encoder. The entropy encoder incorporates a hyperprior model to capture spatial dependencies in the latent representation \cite{balle2018variational}. Meanwhile, the JSCC encoder, built on the MLLA skeleton, compresses the fixed-length latent representation $\bm{z}$ into a variable-length output $\bm{s}$ using the estimated entropy $p(\bm{\tilde{z}}|\bm{\tilde{y}})$. Elements are selected sequentially in a one-dimensional checkerboard pattern to efficiently capture global features.
	
	The receiver comprises a rate-adaptive JSCC decoder and a conditional diffusion decoder. The JSCC decoder reverses the encoder's operations, and its output serves as input to the conditional diffusion decoder, which is built on a U-Net model. Features of different scales, encoded by hierarchical encoders, are combined with time information to guide the denoising process. The conditional diffusion decoder iteratively refines the output until $\bm{\hat{x}}_0$ is produced.
	\subsection{Loss Function}
	For an entropy model-based rate-adaptive JSCC system  \cite{NTSCC}, the loss function is typically defined as:
	\begin{equation} \label{loss_func_rate}
		\begin{aligned}
			\mathcal{L}_{RD}=&\mathcal{D}+\lambda \mathcal{R}
			= \underbrace{\mathbb{E}_{\bm{x}_0}||\bm{x}_0-\hat{\bm{x}}_0||^2}_{\text{JSCC distortion}} + \underbrace{\mathbb{E}_{\bm{x}_0}||\bm{x}_0-\bm{\bar{x}}_0||^2}_{\text{compression distortion}} \\
			+& \lambda \underbrace{\mathbb{E}_{\bm{x}_0}[- \log p(\bm{\tilde{z}}|\bm{\tilde{y}})- \log p(\bm{\tilde{y}})]}_{\text{rate}},
		\end{aligned}
	\end{equation}	
	where $\mathbb{E}_{\bm{x}_0}||\bm{x}_0-\bm{\bar{x}}_0||^2$ is the compression distortion between the source image $\bm{x}_0$ and the compressed image ${\bm{\bar{x}}}_0$, $\mathbb{E}_{\bm{x}_0}[- \log p(\bm{\tilde{z}}|\bm{\tilde{y}})- \log p(\bm{\tilde{y}})]$ represents the compression rate estimated by entropy models, and $\mathbb{E}_{\bm{x}_0}||\bm{x}_0-\hat{\bm{x}}_0||^2$ denotes the typical JSCC loss term.
	
	To improve the perceptual quality of reconstructed images, we introduce an additional perceptual loss, expressed as follows: 
	\begin{equation} \label{loss_func_perception}
		\mathcal{L}_{P} = 
		\mathbb{E}_{\bm{x}_0}[d_p(\bm{x}_0 , \bm{\hat{x}}_0)] + \mathbb{E}_{\bm{x}_0}[d_p(\bm{x}_0,\bm{\bar{x}}_0)],
	\end{equation}	
	where $d_p(\cdot, \cdot)$ represents the perceptual loss term. Here we adopt widely-used learned perceptual image patch similarity (LPIPS) loss based on VGGNet \cite{lpips}. Therefore, the objective for our model can be expressed as:
	\begin{equation} \label{loss_func}
		\begin{aligned}
			\mathcal{L} = &
			(1 - \eta)(\underbrace{\mathbb{E}_{\bm{x}_0}||\bm{x}_0-\bm{\hat{x}}_0||^2}_{\text{JSCC distortion}} + \underbrace{\mathbb{E}_{\bm{x}_0}||\bm{x}_0-\bm{\bar{x}}_0||^2}_{\text{compression distortion}}) \\ 
			+ &
			\eta (\underbrace{\mathbb{E}_{\bm{x}_0}[d_p(\bm{x}_0,\bm{\hat{x}}_0)]}_{\text{JSCC perceptual loss}} + \underbrace{\mathbb{E}_{\bm{x}_0}[d_p(\bm{x}_0,\bm{\bar{x}}_0)]}_{\text{compression perceptual loss}})	\\
			+ & \lambda \underbrace{\mathbb{E}_{\bm{x}_0}[- \log p(\bm{\tilde{z}}|\bm{\tilde{y}})- \log p(\bm{\tilde{y}})]}_{\text{rate}},
		\end{aligned}
	\end{equation}	
	where $\eta \in [0,1]$ balances the trade-off between MSE distortion and the perceptual loss term, while $\lambda$ adjusts the trade-off between the image quality and transmission rate. 
	\subsection{Training Strategy}
	To ensure training stability and enhance overall performance, we propose a multi-stage training strategy. Initially, we train each module individually to reduce complexity and facilitate convergence. Once all modules have been trained separately, we fine-tune the entire model. The complete multi-stage training process is detailed in Algorithm \ref{Training}.

	\begin{algorithm}[!htbp] 
		\caption{Training the proposed CDM-JSCC} 
		\label{Training}
		\SetAlgoLined
		\textbf{Input:} Training dataset $\mathfrak{X}$, trade-off parameters $(\lambda, \eta)$, rate-control parameter $\beta$, and learning rate $l_r$. \\
		\textbf{Output:} Parameters $\left(\bm{\theta}^*_g, \bm{\phi}^*_g, \bm{\theta}^*_h, \bm{\phi}^*_h, \bm{\theta}^*_f,  \bm{\phi}^*_f \right)$. \\
		\textbf{First Stage: Train Compression Modules.}\\
		Randomly initialize all parameters and freeze the parameters of $f_e$ and $f_d$. \\
			\For{each epoch}{
				Sample $\bm{x}$ from $\mathfrak{X}$. \\
				Calculate the loss function:\\
				$\mathcal{L} = 
				(1 - \eta) \mathbb{E}_{\bm{x}_0}||\bm{x}_0-\bm{\bar{x}}_0||^2 
				+ 
				\eta  \mathbb{E}_{\bm{x}_0}[d_p(\bm{x}_0,\bm{\bar{x}}_0)]
				+  \lambda {\mathbb{E}_{\bm{x}_0}[- \log p(\bm{\tilde{z}}|\bm{\tilde{y}})- \log p(\bm{\tilde{y}})]}.$ \\
				Update the parameters $\left(\bm{\theta}_g, \bm{\phi}_g, \bm{\theta}_h, \bm{\phi}_h \right)$. \\
			} 
			------------------------------------------------------------ \\
			\textbf{Second Stage: Train Transmission Modules.}\\
			Load and freeze the parameters trained in the first stage and randomly initialize $f_e$ and $f_d$. \\
			\For{each epoch}{
				Sample $\bm{x}$ from $\mathfrak{X}$. \label{step1}\\
				Calculate the loss function based on (\ref{loss_func}). \label{step2}\\
				Update the parameters $\left(\bm{\theta}_f,  \bm{\phi}_f \right)$.\\
			} 
			------------------------------------------------------------ \\
			\textbf{Third Stage: Finetune the whole model.}\\
			Load the parameters trained in the previous stages. \\
			\For{each epoch}{
				Repeat step \ref{step1} to \ref{step2}. \\
				Update the parameters $\left(\bm{\theta}_g,\bm{\theta}_h, \bm{\phi}_g, \bm{\phi}_h, \bm{\theta}_f,  \bm{\phi}_f \right)$.\\
			} 
		\end{algorithm}

		\section{Simulation Results} \label{S4}
		In this section, we perform simulations to evaluate the performance of the proposed CDM-JSCC.
		\begin{figure}[!t]
			\begin{centering}
				\captionsetup[subfloat]{font=scriptsize}
				\subfloat[The performance on Kodak dataset with average $\text{CBR}=1/48$.]{\label{performance}\includegraphics[width=0.47\textwidth]{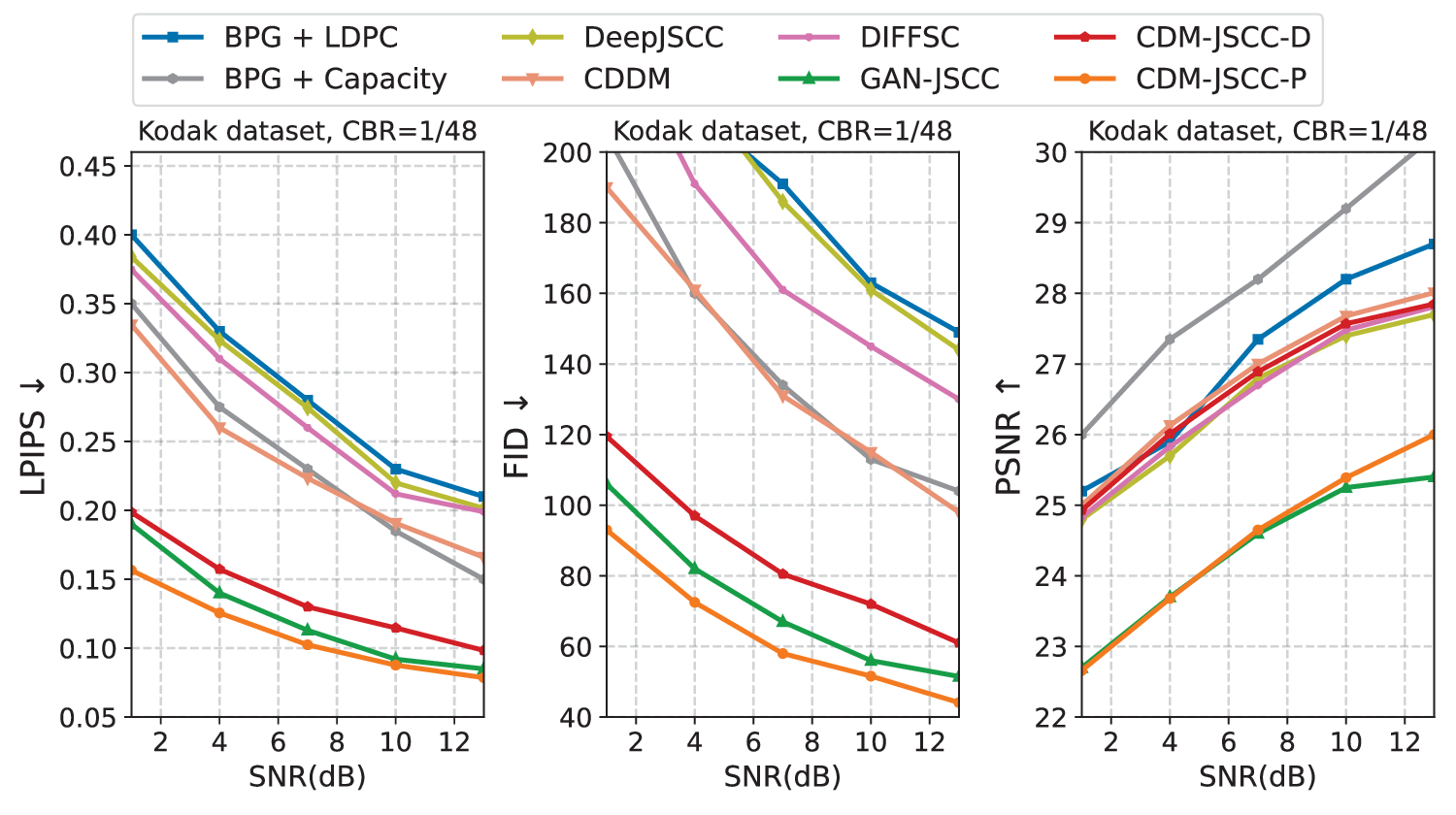}} \\
				\subfloat[The performance on Kodak dataset with average $\text{CBR}=1/24$.]{\label{performance3}\includegraphics[width=0.47\textwidth]{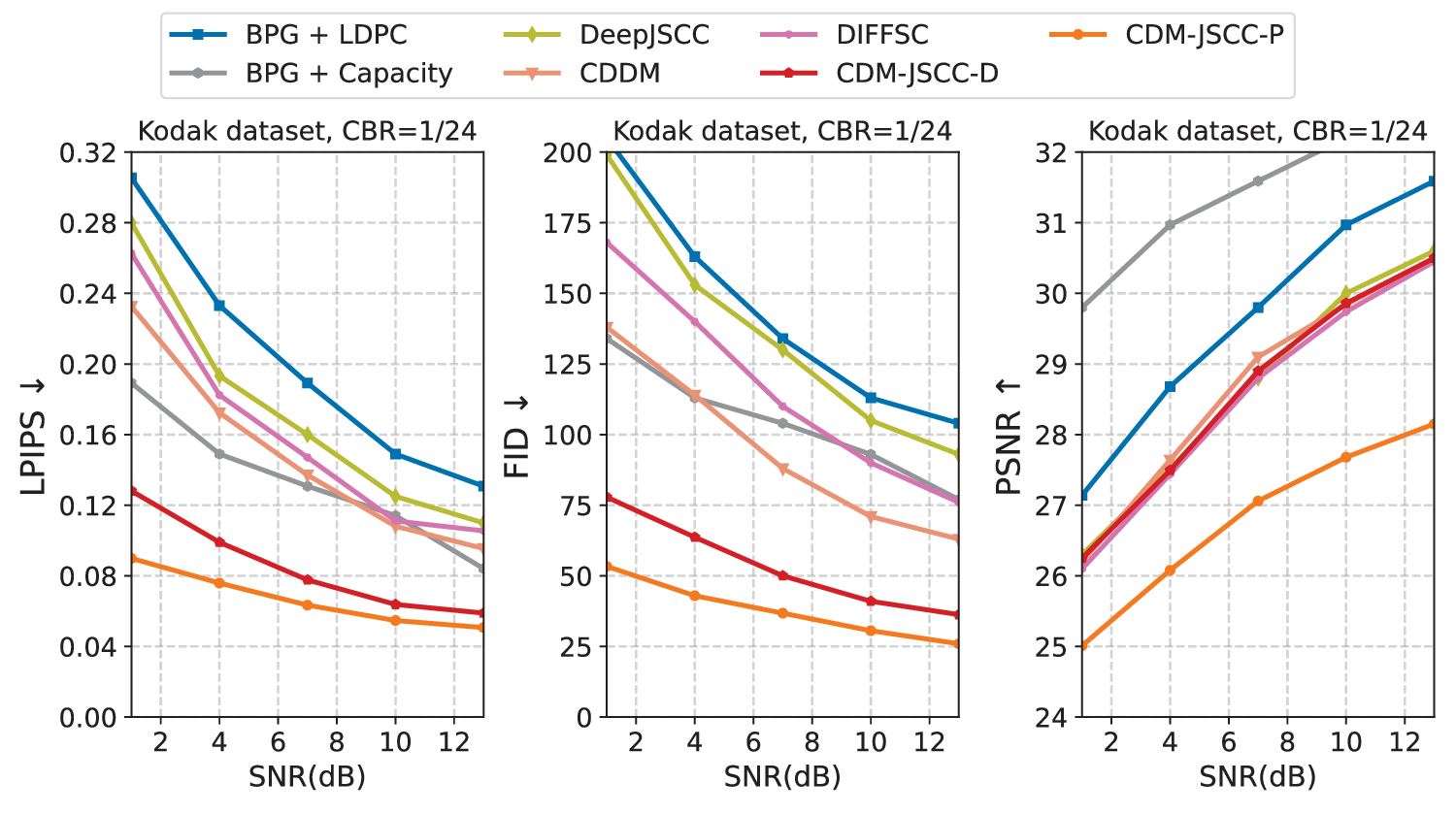}} \\
				\subfloat[The performance on MS-COCO dataset with average $\text{CBR}=1/48$.]{\label{performance2}\includegraphics[width=0.47\textwidth]{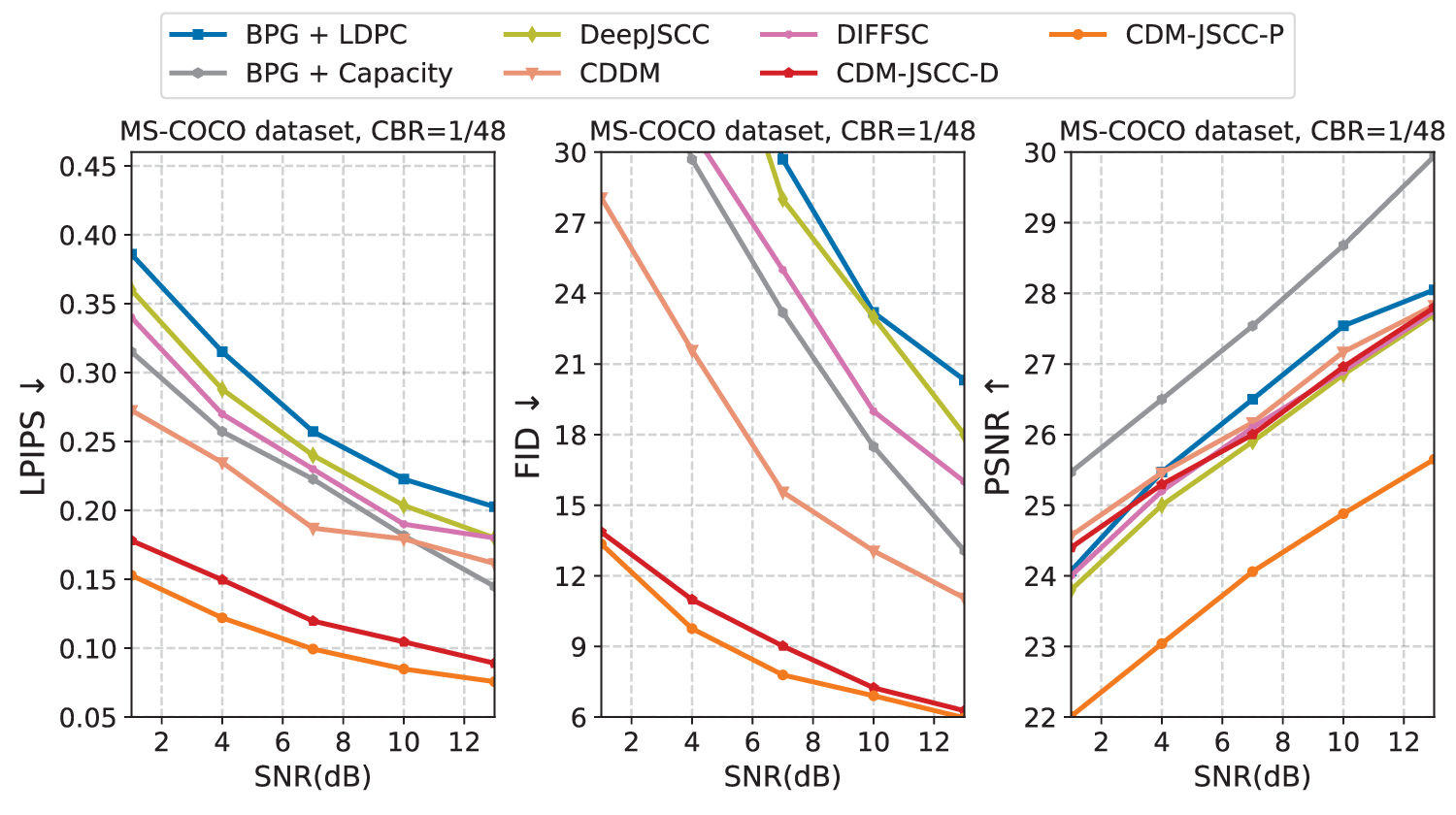}}
				\caption{The performance of the proposed schemes versus SNR.} 
				\label{Performance}
			\end{centering}
		\end{figure}
		\subsection{Simulation Settings}
		\subsubsection{Basic Settings}
		We evaluate our CDM-JSCC on the Kodak dataset ($24$ images, each with a resolution of $512 \times 768$) and MS-COCO dataset (we select $2,695$ test images larger than $512 \times 512$). 
		Our CDM-JSCC trained with $\eta = 0.5$, primarily optimized for high perceptual quality, is referred to as CDM-JSCC-P. In contrast, the model trained with a lower $\eta = 0.1$, aimed at minimizing MSE distortion, is denoted as CDM-JSCC-D. 
		\color{black} All experiments are conducted using Pytorch. For the training process, we use the Adam optimizer for stochastic gradient descent, starting with a learning rate of $1 \times 10^{-4}$ and planning to reduce it after several epochs. The training dataset comprises $50,000$ randomly sampled images from the ImageNet dataset, which are randomly cropped to $256 \times 256$. The batch size is set to $4$, with the epochs for each training stage set to $90, 45$, and $10$, respectively. 
		\color{black}

%
%
		
		\subsubsection{Benchmarks} 
		For benchmarks, we consider both deep learning-based methods and traditional separation-based schemes, incorporating optimization strategies for both MSE distortion and perceptual loss. The benchmarks are detailed as follows:
		\begin{itemize}
			\item ``BPG+LDPC": This method employs BPG for source coding and LDPC for channel coding, followed by quadrature amplitude modulation (QAM).
			\item ``BPG+capacity": In this approach, we use an ideal capacity-achieving channel coding in conjunction with BPG and QAM. 
			\item ``DeepJSCC": As a representative learning-based method, we include the classic DeepJSCC optimized for MSE distortion \cite{JSCC}.
			\item ``GAN-JSCC": This method optimizes both MSE distortion and perceptual loss \cite{wang2022perceptual}.
			\item ``CDDM": This scheme utilizes diffusion models to denoise the channel output \cite{cddm2024wu}.
			\item ``DIFFSC": This method applies diffusion models to enhance images initially reconstructed by DeepJSCC \cite{jiang2024diffsc}.
		\end{itemize}

		\color{black}
		\subsubsection{Considered Metrics}
		We evaluate the performance of our method along with benchmarks using typical metric PSNR as well as perceptual metrics LPIPS\cite{lpips}, and Fréchet Inception Distance (FID). PSNR calculates pixel-wise MSE distortion. LPIPS measures the $l_2$ distance between two latent embeddings extracted by a pre-trained network \cite{lpips}. FID calculates the distribution of source images and reconstructed images in the feature space using the Fréchet distance \cite{FID}.

		\subsection{Performance Comparison}
		
		We evaluate the performance of our proposed model on the AWGN channel, a representative and widely adopted choice in this research community. In all subsequent experiments, the model is trained on a single SNR value and tested on the same SNR.

		Fig. \ref{Performance}(a) compares CDM-JSCC with benchmarks on the Kodak dataset using an average $\text{CBR}=1/48$ across various SNRs, while Fig. \ref{Performance}(b) presents the results at an average $\text{CBR}$ of $1/24$. In addition, Fig. \ref{Performance}(c) shows the results on the MS-COCO dataset at an average $\text{CBR}$ of $1/48$. All results
		demonstrate significant advantages of our proposed CDM-JSCC over the benchmarks, with robust performance across various datasets, CBRs, and SNRs. Notably, our CDM-JSCC-P outperforms all the benchmarks across all perceptual metrics, including LPIPS and FID. This enhancement is attributed to the proposed loss function and the powerful generative diffusion models. However, it is important to note that our CDM-JSCC-P shows performance degradation in the PSNR metric. This is because generative model-based CDM-JSCC-P tends to generate realistic and clear images and may overlook pixel-wise fidelity sometimes. Despite this, our CDM-JSCC-P still outperforms GAN-JSCC in the PSNR metric. Meanwhile, our CDM-JSCC-D is also competitive. Setting $\eta = 0.1$ leads to a larger MSE distortion term compared to $\eta = 0.5$, thereby achieving better pixel-level fidelity measured by PSNR. As shown in Fig. \ref{Performance}, CDM-JSCC-D not only surpasses DeepJSCC and DIFFSC in PSNR but also offers significantly improved perceptual quality over most benchmarks. Besides, the decoding time of CDM-JSCC is approximately $0.76$s per Kodak image, showing advantages over BPG+LDPC of $4.6$s and CDDM of $1.73$s.
		
		\color{black} 
		\section{Conclusion} \label{S5}
		In this letter, we proposed a framework of conditional diffusion models-based generative DJSCC system for image transmission. Moreover, we employed $\mathcal{X}$-prediction with a few denoising steps to accelerate the decoding process. Furthermore, we effectively managed the transmission bandwidth based on the estimated entropy of transmitted symbols. Besides, we proposed a multi-state training strategy to ensure the stability of the training process. Simulation results demonstrated that the proposed method can significantly surpass existing methods in terms of perceptual quality.

		\bibliographystyle{IEEEtran}
		\bibliography{IEEEabrv,Reference}
		
	\end{document}